\documentclass[8pt,letter]{article}
\usepackage{amsmath}
\usepackage{amsfonts}
\usepackage{amssymb}
\usepackage{graphicx}% Include figure files
\usepackage{dcolumn}% Align table columns on decimal point
\usepackage{bm}
\usepackage[utf8]{inputenc}
\usepackage[T1]{fontenc}
\usepackage{mathptmx}
\usepackage{etoolbox}
\usepackage{enumitem}
\usepackage[table]{xcolor}
\usepackage{changes}

\usepackage[pdftex,bookmarks=true,backref=page,pagebackref=false,colorlinks=true, linkcolor=blue]{hyperref}

\usepackage{authblk}
\usepackage{orcidlink}  % Add this package
\usepackage{geometry}

\usepackage{fancyhdr}

\fancyhfoffset[R]{0.0cm}  % Adjust the value to move further right

\newcommand{\exb}{\mathbf{E} \hskip -1pt \times \hskip -1pt \mathbf{B}}
\newcommand{\beunit}{\beta_{e,\mathrm{unit}}}
\newcommand{\bunit}{B_\mathrm{unit}}
\newcommand\dphi{\delta \phi}
\newcommand\dap{\delta \hskip -1pt A_\parallel}
\newcommand\dbp{\delta \hskip -1pt B_\parallel}
\title{Electron-Scale-Driven Turbulence by Negative-Density-Gradient in NSTX}

\author[1]{C F Clauser \orcidlink{0000-0002-2597-5061}\thanks{Corresponding author: cclauser@psfc.mit.edu}}
\author[2]{J Candy \orcidlink{0000-0003-3884-6485}}
\author[3]{J Parisi\orcidlink{0000-0003-1328-7154}}
\author[4]{T Rafiq\orcidlink{0000-0002-2164-1582}}
\author[5]{W Guttenfelder\orcidlink{0000-0001-8181-058X}}
\author[6]{G Avdeeva\orcidlink{0000-0001-7072-7967}}
\author[2]{E Belli\orcidlink{0000-0001-7947-2841}}
\author[7]{J Berkery\orcidlink{0000-0002-8062-3210}} 
\author[8]{I Sfiligoi\orcidlink{0000-0002-9308-5327}}
\author[4]{E Schuster\orcidlink{0000-0001-7703-6771}}
        
\affil[1]{Massachusetts Institute of Technology, Cambridge MA 02139,  USA}
\affil[2]{General Atomics,  La Jolla CA 92121,  USA}
\affil[3]{Marathon Fusion,  San Francisco CA 94107,  USA}
\affil[4]{Lehigh University,  Bethlehem PA 18015,  USA}
\affil[5]{Type One Energy, Oak Ridge TN 37830, USA}
\affil[6]{Prism Computational Science, Inc, Madison WI 53711, USA}
\affil[7]{Princeton Plasma Physics Laboratory,  Princeton NJ 08540,  USA}
\affil[8]{University of California San Diego, La Jolla CA 92093,  USA}

\begin{document}

\maketitle

\begin{abstract}
Gyrokinetic simulations of an NSTX spherical-tokamak plasma reveal novel negative-density-gradient (NDG) drift waves at electron-gyroradius-scale that drive turbulent transport comparable to the experimental power flow. Linear simulations indicate that the dominant instability is a trapped-electron electron-scale tearing-parity mode driven mainly by negative electron density gradient, $a/L_{ne} \equiv - a/n_e \, (dn_e/dr) < 0$. Although thermal transport is mainly carried by transverse magnetic, $\dap$, fluctuations, the growth rate is sensitive to compressional magnetic fluctuations, $\dbp$, differentiating these from standard microtearing modes (MTMs). Linear sensitivity analysis also reveals that the modes are most unstable at lower collisionality, consistent with the trapped-electron character. Electron-scale and multi-scale nonlinear simulations show that these modes can account for several megawatts of experimental power, with the latter suggesting that experimental gradients may lie at a bifurcation between distinct turbulence regimes.
\end{abstract}

% Now switch to two columns with reduced margins
\newgeometry{
    top=1.5cm,
    bottom=1.5cm,
    left=1.5cm,
    right=1.5cm,
    columnsep=1cm      % Space between columns
}
\setlength{\columnsep}{1cm}  % Increase column separation

\twocolumn
\sloppy
%\begin{multicols}{2}

\noindent \textit{Introduction. }
Repeated NSTX tokamak plasma experiments have demonstrated that the ion thermal transport is often neoclassical, subdominant to turbulent electron thermal transport \cite{Kaye2007,Kaye2007a}. In particular, microtearing modes (MTMs) and electron temperature gradient modes (ETG) have been identified as likely candidates for anomalous heat transport \cite{Guttenfelder2011,Guttenfelder2013,Clauser2025,Belli2025}. MTMs are typically found at ion-gyroradius scales ($k_{\theta} \rho_s < 1$) \cite{Guttenfelder2011,Drake1980}, although in high $\beta$ spherical tokamaks (STs), they have been identified at $k_{\theta} \rho_s > 1$ near the magnetic axis in NSTX \cite{Smith2011} and for pilot-plant designs \cite{Patel2022}. Here, $k_{\theta}=nq/r$, which relates to the bi-normal wave number $k_y$ ($n$ is the toroidal wave number, $q$ is the safety factor and $r$ the radial coordinate), $\rho_s=c_s (m_D c)/(e\bunit)$ is the ion-sound gyroradius, $c_s = \sqrt{T_e/m_D}$ is the deuterium sound speed evaluated at the electron temperature and $\bunit=(q/r)d\psi/dr$ the effective magnetic field (see details in \cite{Candy2016}).

NSTX plasmas usually exhibit dynamic density profiles in which negative density gradients, defined here as $a/L_n \equiv -a/n \, (dn/dr) < 0$, where $a$ is the minor radius, develop over the course of the discharge, usually as a consequence of a low-to-high confinement (L-H) mode transition. Such gradients are typically stabilizing for both MTM and ETG instabilities. Microinstabilities in negative-density-gradient regions have been investigated in different devices, particularly in the context of pellet or neutral beam injection (NBI). Transport due to hollow (or negative-gradient) density profiles has been studied using gyrokinetic codes in JET NBI plasmas \cite{Baiocchi2015} with focus on particle transport, in pellet injection in MAST \cite{Garzotti2014} with analysis limited to linear modes, and in ASDEX-U \cite{Angioni2017} with the analysis limited to electrostatic ion-scale modes. None of these prior studies reported significant thermal transport due to negative-density-gradient (NDG) driven modes.

In this Letter, we describe novel NDG trapped-electron electron-scale tearing-parity modes that drive transport comparable to estimated power-balance values. The numerical analysis is based on fully-electromagnetic $(\dphi,\dap,\dbp)$ linear and nonlinear CGYRO \cite{Candy2016,candy:2019} simulations. Both electron-scale and preliminary multiscale nonlinear simulations are reported here. Baseline parameters are based on NSTX discharge $\#120982$ (TRANSP ID 120982A09) at $620\,$ms. This discharge was used in previous studies \cite{Guttenfelder2013,Clauser2022}.
\begin{figure}[ht]
\vspace{-1mm}
\centering
\includegraphics[width=0.95\columnwidth]{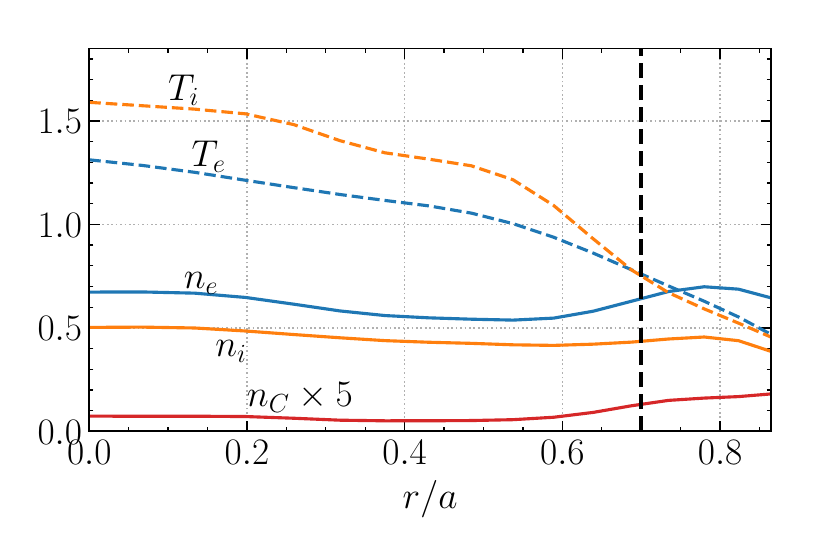}
\vspace{-7mm}
\caption{Temperature (keV) and density ($10^{20}\,\text{m}^{-3}$) profiles for both electrons and the main ion species (deuterium), and impurity carbon density profile. We study the flux surface $r/a=0.7$, marked with a vertical dashed line.}
\label{fig:1}
\vspace{-1mm}
\end{figure}

Figure \ref{fig:1} shows temperature and density profiles for both electrons and main ion species (deuterium), as well as the density of carbon. The radial location of interest in this work is $r/a=0.7$, indicated with the dashed vertical line.  At this radial location, $(a/c_s) \gamma_E \sim 0.21$, where $\gamma_E = (r/q) d\omega_0/dr$ is the Waltz $\exb$ shearing rate \cite{Waltz1999}, with $\omega_0(r)$ the toroidal rotation frequency \cite{belli:2018}. The shearing rate cannot be measured directly but must be calculated using neoclassical force balance arguments and for this reason we consider its uncertainty to be high as detailed in Sec.~IIA of Ref.~\cite{avdeeva:2025}. Frequencies are measured in units of $c_s/a$. Local parameters at this radius are summarized in Table~\ref{tab.local}. All simulations were conducted using three kinetic species: thermal electron, thermal deuterium, and carbon (as the main impurity). Because the plasma contains a small amount of hydrogen in addition to deuterium, we consider the main ion mass to be $m_i = 0.95 m_D$. Additional parameters can be found in Refs. \cite{Clauser2025,Clauser2022}.
\begin{table}
\caption{Baseline parameters at $r/a=0.7$. The flux-surface shape is exactly up-down symmetric. The elongation $\kappa$ and triangularity $\delta$, along with their gradients $s_\kappa$ and $s_\delta$, are defined in \cite{arbon:2021}. Here, $\Delta = d R_0/dr$ is the Shafranov shift.}
\begin{center}
\begin{tabular}{|c|c||c|c|}
\hline
Parameter           & Value  & Parameter     & Value \\
\hline
$R_0/r$             & 1.56   & $\Delta$      & -0.4 \\
$q$                 & 3.5    & $s$           & 1.36 \\
$\kappa$            & 2.2    & $s_\kappa$    & 0.004 \\
$\delta$            & 0.28   & $s_\delta$    & 0.27 \\
$(a/c_s)\gamma_E$   & 0.21   & $T_i/T_e$     & 1.0  \\
$(a/c_s) \nu_{ee}$  & 0.5    & $\beunit$     & 0.93\% \\
$a/L_{Ti}$          & 3.3    & $a/L_{Te}$    & 2.2 \\
\hline
\end{tabular}
\end{center}
\label{tab.local}
\end{table}  
\begin{table}
\caption{Table of quasineutral density gradients. Although the best-estimate for the experimental value of $a/L_{n_e}=-1.53$, we use $a/L_{n_e}=-1.6$ as the baseline experimental value (see grey shaded row). For linear sensitivity analysis (see Fig. \ref{fig:lin03})  we use the gradients in the blue shaded row.}
\begin{center}
\begin{tabular}{|c|c|c|}
\hline
$a/L_{ni}$ & $a/L_{nC}$ & $a/L_{ne}$ \\
\hline
\rowcolor{gray!20}
-0.66 & -4.6 & -1.6 \\
-0.93 & -4.6 & -1.8 \\
\rowcolor{blue!20}
-1.19 & -4.6 & -2.0 \\
-1.46 & -4.6 & -2.2 \\
-1.73 & -4.6 & -2.4 \\
\hline
\end{tabular}
\end{center}
\label{tab.scan}
\end{table}  
\begin{figure}
\vspace{-1mm}
\centering
\includegraphics[width=0.98\columnwidth]{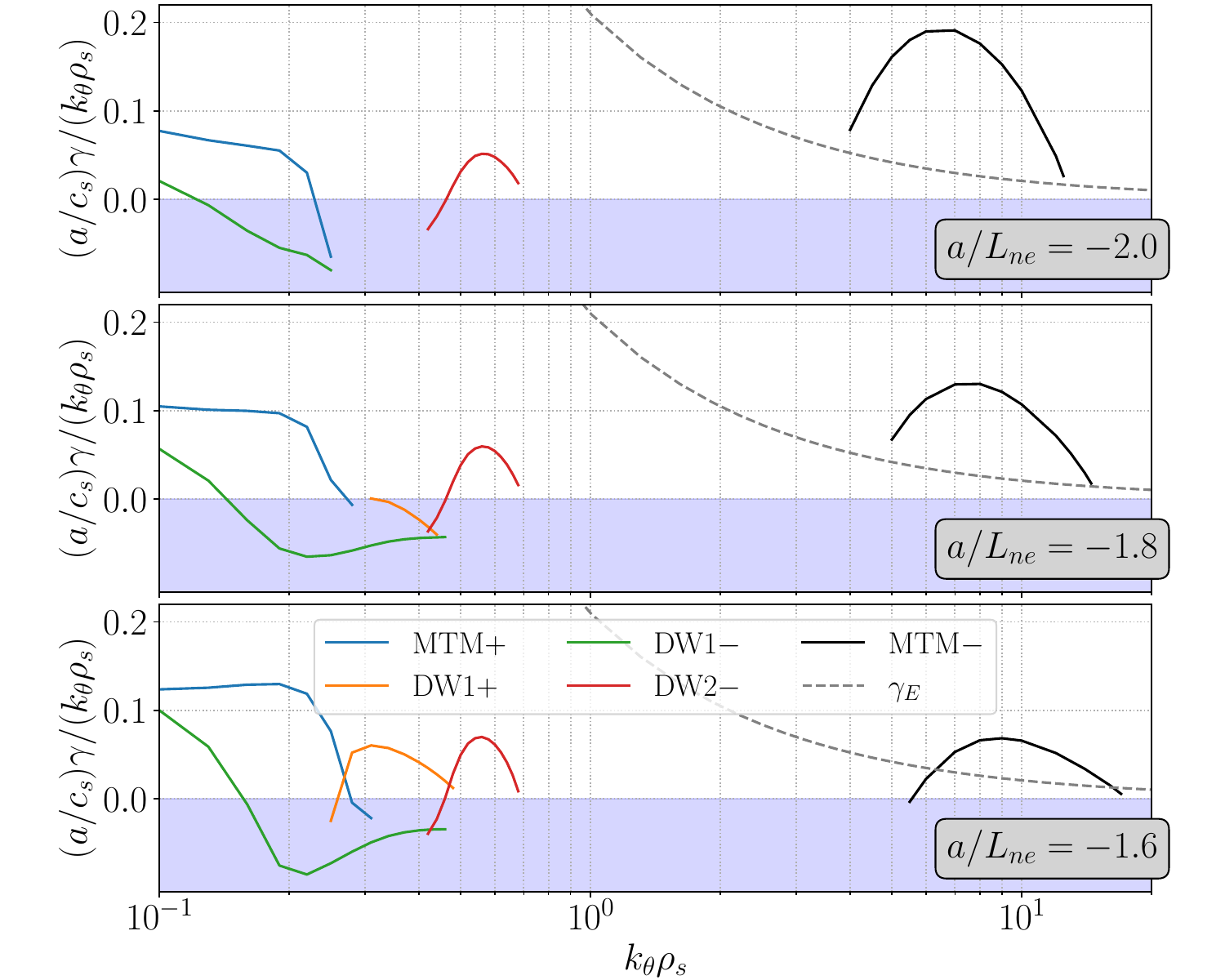}
\vspace{-3mm}
\caption{Normalized growth rate versus wavenumber $k_{\theta}\rho_s$ for three values of $a/L_{n_e}$. At short wavelength, we observe a single unstable MTM (which will be referred later as NDG), whereas at long wavelength we see at least four different drift waves (DW): one MTM (blue), two negative-frequency (ion-direction) drift waves (green, red) and one positive-frequency (electron-direction) drift wave (orange). Notably, as $a/L_{n_e}$ increases in magnitude from $-1.6$ to $-2.0$, the growth rate of the electron-scale MTM grows while the growth rate of the four ion-scale modes decreases. Flow shear, $(a/c_s)\gamma_E=0.21$, is indicated with dashed lines.}
\label{fig:lin01}
\vspace{-3mm}
\end{figure}

\vspace{2mm}
\noindent \textit{Linear analysis. }
Linear analysis was carried out with both the initial-value CGYRO solver and the newer CGYRO-DMD eigenvalue solver \cite{dudkovskaia:2025}. The latter computes subdominant modes in addition to the most unstable mode. Fully electromagnetic (3-field) perturbations were evolved over the full range of unstable wavenumbers and for three values of $a/L_{n_e}$. Table \ref{tab.scan} shows density gradients for the kinetic species employed in this work. Although the best-estimate for the density gradient experimental value is $a/L_{n_e}=-1.53$, for simplicity, we use $a/L_{n_e}=-1.6$ as the baseline value.

In Fig.~\ref{fig:lin01} we plot linear growth rates $(a/c_s)\gamma$ normalized to wavenumber $k_\theta \rho_s$, for three different density gradients: $-1.6$ $-1.8$ and $-2.0$. The normalization helps to better emphasize the long-wavelength part of the spectrum (so called ion-scale in which $k_\theta \rho_s \leq 1$) which typically dominates the transport even though the short wavelength growth rates (so called electron-scale in which $k_\theta \rho_s > 1$) may be much larger. The results have a number of notable features. First, there seems to be a clear separation between unstable modes in the ion scale and electron scale, with no unstable modes between $0.7<k_\theta \rho_s < 5.0$ for the nominal case. At short wavelength, we observe only a single unstable tearing-parity mode, which we label for now as $\mathrm{MTM-}$ (it will be referred to later as an NDG mode). The sign here indicates that the mode has a negative frequency, so it is in the ion-direction. At long wavelength, we see at least 4 different drift waves: one electron-direction mode labeled $\mathrm{MTM+}$, two ion-direction drift waves $\mathrm{DW-}$ and one electron-direction drift wave $\mathrm{DW+}$. Some of the ion-scale modes were previously identified in Ref. \cite{Clauser2022}. 
To assess the direction of the short-wavelength mode $\mathrm{MTM-}$, we consider the electron-scale limit $k_\perp \rho_s \gg 1$ with adiabatic ions for which the linear gyrokinetic dispersion relation reduces to a simple electron balance. Taking $|\omega| \gg |k_\parallel v_{te}|,|\omega_{de}|$, the leading-order solvability condition gives a real frequency proportional to the electron density gradient. Neglecting electron FLR gives
\[
(a/c_s) \omega \sim k_\theta \rho_s \, \frac{a/L_n}{2-T_e/T_i} \; .
\]
When $a/L_n < 0$, this limit predicts an ion-direction mode.
As $a/L_{n_e}$ increases in magnitude from $-1.6$ to $-2.0$ (a $25\%$ increase), the growth rate of the electron-scale $\mathrm{MTM-}$ doubles, while the growth rates of all ion-scale modes decrease (or remain unchanged for DW$-$). A simple extrapolation of the growth rates, allow us to infer that the MTM$-$ exhibits a threshold at $a/L_{ne} \approx -1.45$, therefore being unstable at the nominal experimental value. This highlights the strong sensitivity of these modes to negative density gradients. Note that the actual peak growth rate the $\mathrm{MTM-}$ modes (un-normalized by $k_{\theta}\rho_s$) is between one to two orders of magnitude larger than the ion-scale modes. Further, the growth rates of all ion-scale modes are less than the nominal value of the $\exb$ flow shear, $(a/c_s)\gamma_E=0.21$, suggesting that ion-scale turbulence may be partially or fully suppressed. 
\begin{figure}[ht]
\vspace{-1mm}
\centering
\includegraphics[width=0.95\columnwidth]{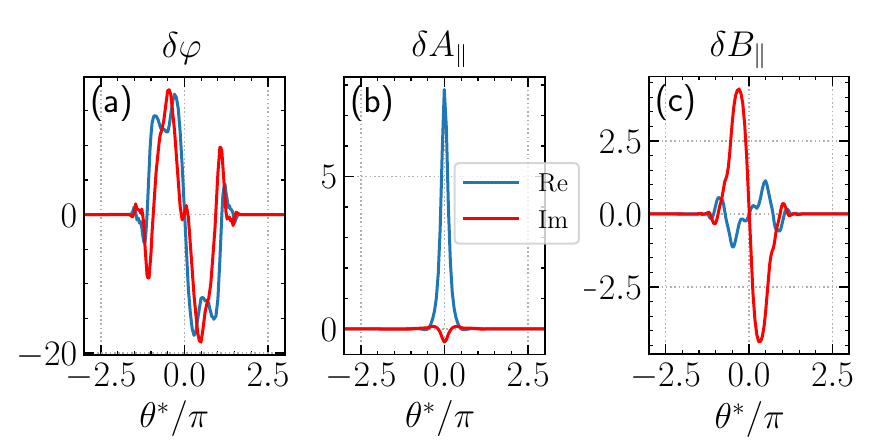}
\vspace{-3mm}
\caption{Eigenfunctions of the normalized electrostatic (a), transverse magnetic (b) and compressional magnetic (c) potential are plotted as functions of the extended angle $\theta_*$ for $k_{\theta}\rho_s = 10$ at nominal experimental conditions. The mode exhibits tearing parity.}
\label{fig:lin02}
\vspace{-3mm}
\end{figure}

Fig. \ref{fig:lin02} shows curves of the normalized electrostatic potential, transverse magnetic and compressional magnetic fields, $e \dphi / T_e$, $(c_s/c)\, e \dap/T_e$, $\dbp/B_\mathrm{unit}$, respectively, at $k_{\theta}\rho_s = 10$ for the baseline parameters. These quantities are plotted as functions of the extended angle $\theta_*$.  While this mode has tearing parity, without including $\dbp$ the growth rate is drastically reduced. This is in contrast to classic (or collisional) MTMs for which $\dbp$ fluctuations are relatively unimportant. It is worth noting that inclusion of $\dbp$ has recently been seen to be important for NSTX simulations of ion temperature gradient and trapped electron modes \cite{Kinsey2025}, ETG modes \cite{Clauser2025,Li2025}, and altering mode structures from MTM to kinetic ballooning modes \cite{McClenaghan2025}. This mode, like any tearing-parity mode, undergoes magnetic reconnection yet critically requires $\dbp$ for drive, meaning that perpendicular pressure dynamics are essential for the mode character.  In addition, quasilinear flux ratios obtained from linear simulations show that the main transport channel is the electron thermal transport, having $Q_i/Q_e \sim 10^{-4}$, and that this is dominated by the $\dap$ contribution. Here, $Q_i$ and $Q_e$ are the ion and electron heat fluxes, respectively. We have verified that the mode is unstable only in a narrow range around $\theta_0=0$, the ballooning angle corresponding to the most unstable core-plasma eigenmode (typically). 

\begin{figure}[ht]
\vspace{1mm}
\centering
\includegraphics[width=0.98\columnwidth]{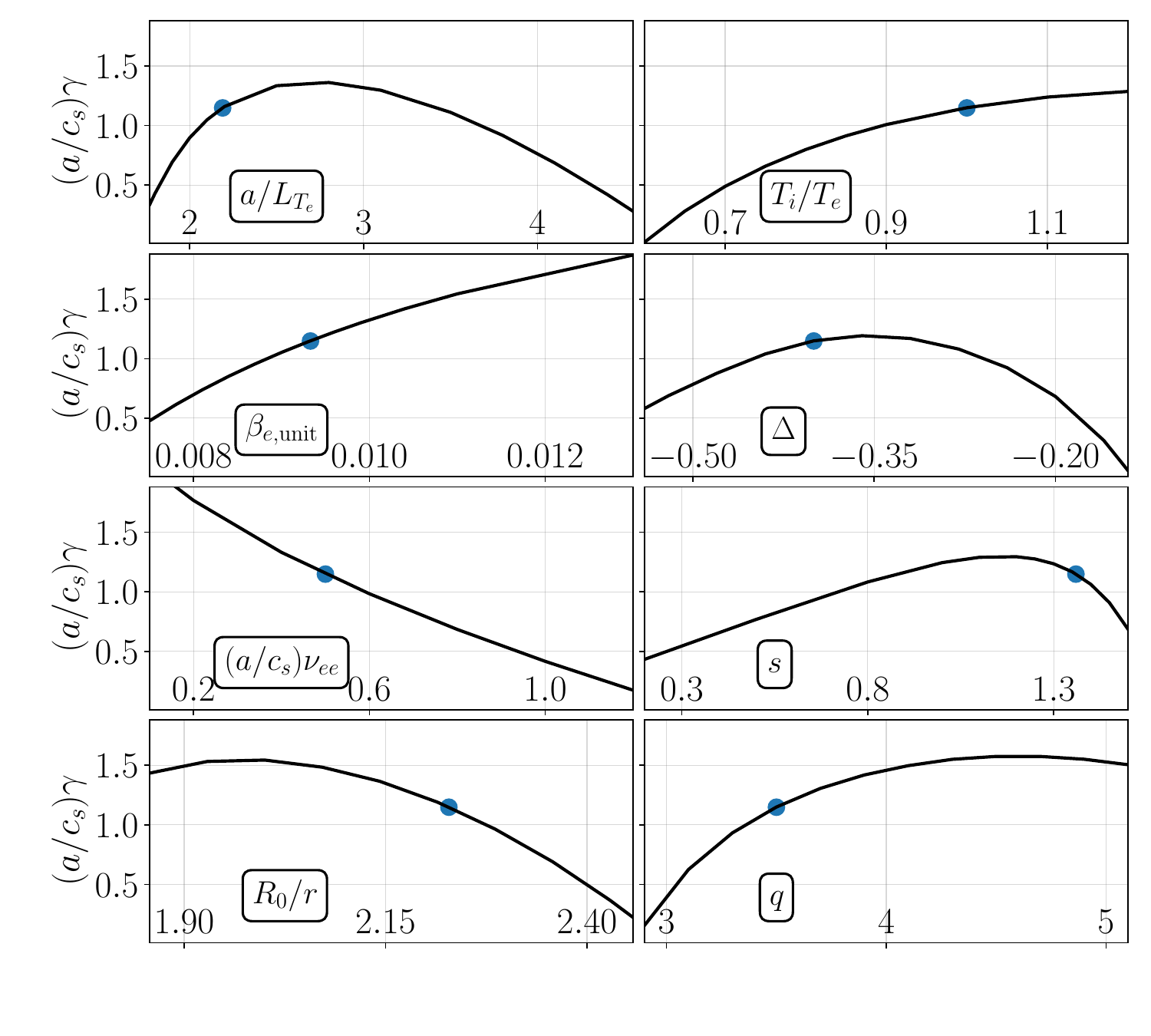}
\vspace{-8mm}
\caption{Mode sensitivity analysis showing growth rates for the $k_{\theta}\rho_s = 10$ scanned over electron temperature gradient $a/L_{T_e}$, temperature ratio $T_i/T_e$, effective electron plasma beta $\beunit$, Shafranov shift $\Delta$, electron collision rate $(a/c_s)\nu_{ee}$, shear $s$, aspect ratio $R_0/r$ and safety factor $q$. Blue dot refers to the reference point for the sensitivity analysis (see Table~\ref{tab.scan}).}
\label{fig:lin03}
\vspace{-2mm}
\end{figure}

To more fully clarify the mode character, a sensitivity analysis was carried out at $k_\theta \rho_s = 10$ and at $a/L_{ne}=-2.0$ (as indicated in blue shaded row of Table \ref{tab.scan}), and is summarized in Fig.~\ref{fig:lin03}. This choice was made to emphazise the mode character, but similar conclusions are obtained with $a/L_{ne}=-1.6$. Scans were performed over the electron temperature gradient, $a/L_{T_e}$, ion-electron temperature ratio, $T_i/T_e$, electron plasma beta, $\beunit$, Shafranov shift, $\Delta$, collisionality, $\nu_{ee}$, magnetic shear, $s$, aspect ratio, $R_0/r$, and safety factor, $q$. The blue dot indicates the nominal value. This analysis, in conjunction with the results of Fig.~\ref{fig:lin01}, reveals a clear density-gradient drive and an intrinsically trapped-electron character. The mode exhibits a finite-$\beta$ onset  ($\beunit \approx 0.6\%$) that can be inferred from a simple extrapolation of the $\beunit$ scan and a growth rate that increases strongly with this parameter. On the other hand, it remains comparatively insensitive to electron temperature-gradient drive; specifically, it persists over a broad window in $a/L_{T_e}$ ($1.5 < a/L_{T_e} < 5$) with a moderate and non-monotonic variation across this range (note that the $\beunit$ scanned range was $\sim \pm 30\%$ the nominal value, while $a/L_{T_e}$ was increased more than $100\%$ its nominal value). Trapping physics is essential to the existence of this mode, which vanishes when the electron trapping term in CGYRO (Eq.~(59) of \cite{Candy2016}) is removed. The growth rate decreases monotonically as the electron collisionality $\nu_{ee}$ is increased, consistent with detrapping of trapped electrons. This property suggests that these modes may be present in future NSTX-U plasmas \cite{Guttenfelder2022,Berkery2024,Munaretto2026} as well as in reactor relevant conditions where lower collisionality regimes are expected. The mode is insensitive to safety factor over the range $4 \le q \le 5$, but is stabilized for $q < 3$, indicating a threshold rather than strong continuous $q$-scaling. Similarly, it is weakly dependent on magnetic shear and Shafranov shift over the range $-0.7 < \Delta < -0.15$, and exhibits a broad plateau in aspect ratio, with $\gamma$ nearly unchanged for $1.8 < R/r < 2.2$ but stabilized for sufficiently large aspect ratio ($R/r > 2.5$). 

The large growth rate at vanishing $\nu_{ee}$ shows that collisions are not required to generate the nonadiabatic electron response that sustains the tearing-parity instability. Instead, trapped electrons are responsible for an intrinsically nonadiabatic response even in the collisionless limit. This contrasts with the classic collisional microtearing branch, where the dominant drive mechanism relies on electron collisions to produce the required nonadiabatic parallel current response, leading to a peak in growth rate at intermediate $\nu_{ee}$. Taken together, the following properties support identification as a \emph{negative-density-gradient-driven, trapped-electron microtearing mode}:
\begin{itemize}[leftmargin=15pt]\setlength{\itemsep}{0pt} 
    \item Tearing parity and a finite-$\beta$ threshold, with strong electromagnetic amplification as $\beta$ increases;
    \item Clear density-gradient drive with weak sensitivity to $a/L_{T_e}$;
    \item An intrinsically trapped-electron character;
    \item Monotonic stabilization with increasing $\nu_{ee}$.
\end{itemize}
For simplicity, we name these modes as negative-density-gradient or \textit{NDG} modes.

As discussed above, these modes are present over a wide range of the scanned parameter space, which would well cover experimental uncertainties. Also, as shown in Fig. \ref{fig:lin01} these NDG are the dominant modes with growth rates well above the $\exb$ flow shear rate, while the ion-scale modes' growth rates remain below and further suppressed at increased density gradients. Therefore, it is important to assess whether these NDG modes can produce enough transport that could account for, or contribute significantly to, experimental values. However, it is important to note that, although previous studies in NSTX plasma have shown MTM suppression due to $\exb$ flow shear \cite{Guttenfelder2011}, MTMs can be sometimes resilient to it \cite{Patel2022,Giacomin2023}, which will require multiscale nonlinear simulations to cover the full spectrum.

%\section{Nonlinear simulations}
\vspace{2mm}
\noindent \textit{Nonlinear simulations. }
Power balance from TRANSP indicates that the total experimental power flow through the $r/a=0.7$ flux surface is $\sim3\,$MW (which corresponds to energy flux of $0.11\,\textrm{MW/m}^2$). To explore whether NDG turbulence can produce electron energy transport comparable to experimental levels, we conducted a series of nonlinear simulations. The preceding linear analysis showed clear scale separation between the electron-scale NDG modes and the ion-scale modes, with the latter having growth rates well below $\gamma_E$. This suggested that electron-scale simulations alone might be sufficient, owing to strong suppression of the ion-scale modes, as in the ETG/MTM cases reported in Ref. \cite{Belli2025}. Electron-scale simulations were therefore compared against more accurate multiscale simulations.

\begin{figure}[h!]
%\vspace{0mm}
\centering
\includegraphics[width=0.95\columnwidth]{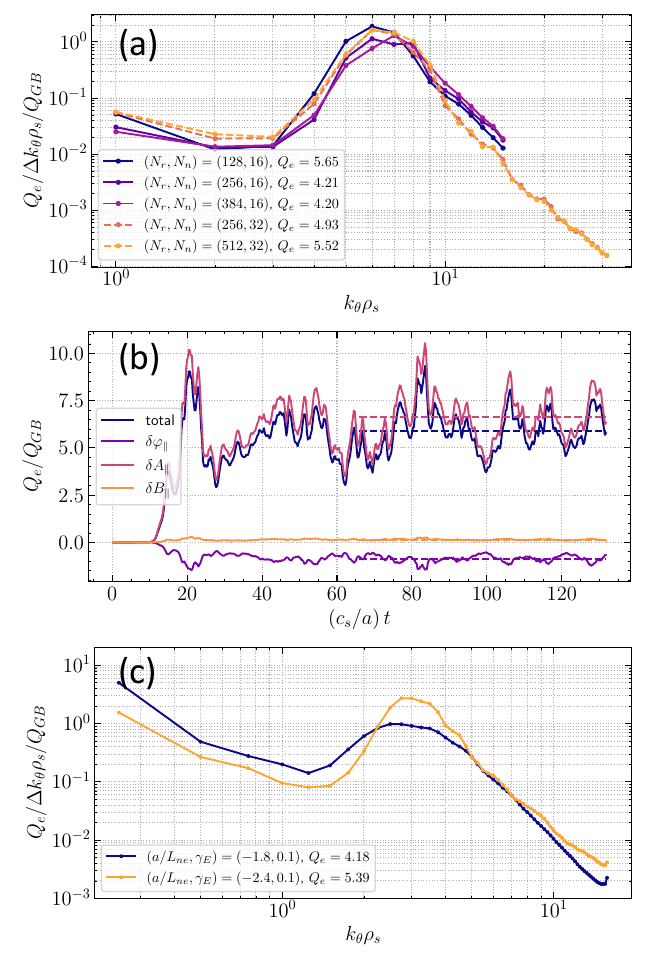}
\vspace{-5mm}
\caption{(a) Electron scale thermal heat flux spectra for different radial ($N_r$) and toroidal ($N_n$) resolutions, at $a/L_{ne}=-1.8$. Values of the averaged electron thermal transport, $Q_e$ (in GB units) is also included for reference. (b) Time evolution of the thermal heat flux for the case $(N_r,N_n)=(128,16)$, showing saturation and contribution from each field. (c) multi-scale thermal heat flux spectra for two density gradients, $a/L_{ne}$.}
\label{fig:nl01}
%\vspace{-3mm}
\end{figure}

All nonlinear simulations used $(a/c_s)\gamma_E=0.1$ to avoid the region of subcritical turbulence. Typically, electron-scale simulations are mostly insensitive to $\gamma_E$ because they do not contain low-$k$ modes that are strongly influenced by flow shear. To begin, we conducted electron-scale convergence tests with $(k_{\theta}\rho_s)_\mathrm{min}=1.0$ and $(k_{x}\rho_s)_\mathrm{min}=0.712$, giving box sizes, $L/\rho_s$, of $6.28$ and $8.82$, respectively. The analysis was also conducted for $a/L_{ne}=-1.8$ to again avoid the region of subcritical turbulence. Figure \ref{fig:nl01}(a) shows a summary of these results. The test was conducted varying toroidal ($N_n$) and radial ($N_r$) resolution, and as can be observed, there is a good agreement among the different cases, even at the lowest resolution $(N_r,N_n)=(128,16)$. The remaining resolutions were $n_\theta=32$ grid points, $n_\xi=12$ pitch-angle Legendre polynomials, and $n_\varepsilon=8$ energy Steen polynomials. The spectral peaks around $k_{\theta}\rho_s=6$ are consistent with NDG modes. Figure \ref{fig:nl01}(b) presents the time evolution of the flux for the lowest resolution case, where contributions from each field are included. It is clear that the main contribution is due to $\dap$. Curiously, the $\dphi$ fluctuations produce a weak inward flux, which might be a consequence of the negative density condition.

To compare with the power balance estimate (see beginning of this section), the energy flux, $Q_e$, is also indicated for convenience in Fig. \ref{fig:nl01}(a). Note that the gyroBohm unit of flux is $Q_{GB}=n_e T_e c_s \rho_s^2/a^2=0.0316\,\textrm{MW/m}^2$, so that, as an example, $Q_e=0.179\,\textrm{MW/m}^2$ for the lowest resolution case. 
Notably, these fluxes significantly exceed the power-balance estimates, indicating that NDG modes can readily account for the experimentally inferred transport. However, despite the linear scale separation and the fact that the ion-scale growth rates are well below $\gamma_E$, electron-scale-only nonlinear simulations exclude low-$k$ couplings by construction. Those couplings, even to linearly stable fluctuations, can influence zonal-flow dynamics and thus modify the saturation level and spectral distribution of the turbulence. 

To explore the effect of low-$k$ coupling, high-resolution multiscale simulations were conducted. We increased toroidal and radial resolution to $N_n=64$ and $N_r=512$ but otherwise used the electron-scale resolution settings. This increase corresponds to $(k_{\theta}\rho_s)_\mathrm{min}=0.25$ and $(k_{x}\rho_s)_\mathrm{min}=0.178$ with box sizes $L/\rho_s$ of $25.13$ and $35.29$, respectively. The maximum wavenumbers were $(k_{\theta}\rho_s)_\mathrm{max}=15.750$ and $(k_{x}\rho_s)_\mathrm{max}=45.396$, which are identical to the values in the electron-scale runs at $(N_r,N_n)=(128,16)$. Even with this choice, a single multi-scale simulation used about $10-12$k node-hrs in the OLCF Frontier supercomputer, drastically limiting the ability to scan over different parameters of interest. Electron energy flux spectra are shown in Fig. \ref{fig:nl01}(c) for two density gradients: $a/L_{ne}=-1.8$ and $-2.4$. At $a/L_{ne}=-2.4$, the NDG wavenumber range dominates the spectrum, whereas at $a/L_{ne}=-1.8$ the NDG peak is weakened and the turbulence begins to shift to the low-wavenumber range suggesting a transition to ion-scale turbulence.

\begin{figure}[h!]
%\vspace{0mm}
\centering
\includegraphics[width=1.0\columnwidth]{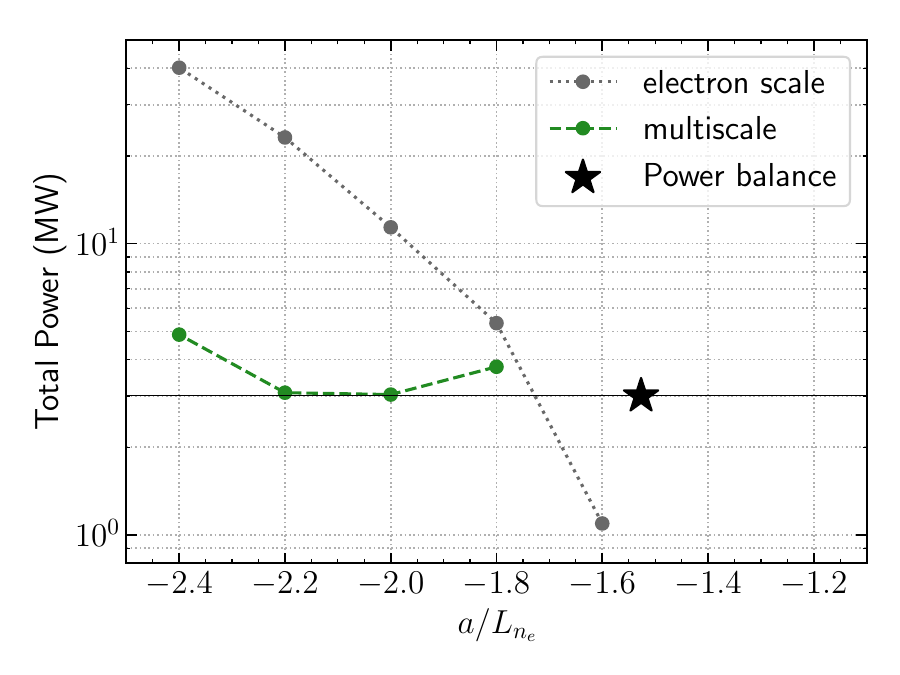}
\vspace{-9mm}
\caption{Power flow at the $r/a=0.7$ flux surface. Nonlinear electron-scale and multiscale simulations were conducted varying the density gradient, $a/L_{ne}$. The star shows the value from TRANSP power balance.}
\label{fig:nl03}
%\vspace{-3mm}
\end{figure}

To better assess the electron thermal transport caused by these modes, we completed a scan over the electron density gradient $a/L_{ne}$. The summary of these simulations is presented in Fig. \ref{fig:nl03}, which shows the total power flow (note log scale) versus $a/L_{ne}$. The power balance value calculated with TRANSP is also included for reference.

In the electron-scale scan, as $|a/L_{ne}|$ increases the power flow greatly exceeds that of the power balance, whereas as $|a/L_{ne}|$ is reduced, the NDG transport is suppressed. The multiscale simulations however show a more complex feature: the flux appears to transition from electron-scale NDG turbulence at large $|a/L_{ne}|$ ion-scale turbulence at reduced $|a/L_{ne}|$. Therefore, the \textit{separability hypothesis}, whereby one assumes that electron-scale simulations accurately represent the electron-scale spectrum applies only marginally to these plasma parameters. This result aligns with the growing evidence that the experimental parameters may lie at a bifurcation of turbulent regimes (i.e. between ion and electron scale) \cite{Belli2023}.

\vspace{2mm}
\noindent \textit{Conclusions. }
For the first time, linear and nonlinear gyrokinetic simulations reveal a novel electron-scale trapped-electron microtearing mode driven by negative density gradients, which can produce thermal transport levels that can significantly contribute to those inferred experimentally. These modes are referred here as NDG modes. 
The modes also exhibit electromagnetic transport dominated by $\dap$, with sensitivity to compressional magnetic fluctuations, $\dbp$, and a finite threshold and strong sensitivity to plasma $\beta$. A collisionality scan suggests they would be relevant in ST regimes, consistent with their trapped-electron nature.
Comparisons between nonlinear electron-scale and multiscale simulations indicate that electron-scale simulations reproduce some qualitative aspects of the transport behavior, but that the spectral shape and total energy flux may differ from more realistic multiscale values.
Further work is required to better understand the occurrence of these modes in ST devices where negative density gradients are present. In particular, pellet injection fueling can generate strong negative density gradient regions and may trigger NDG modes \cite{Garzotti2014}. Additional studies are also needed to determine the broader importance of these modes for electron heat transport in tokamak plasmas.

%\begin{acknowledgments}
\vspace{2mm}
\noindent \textit{Acknowledgments. }
This work was supported by the U.S. Department of Energy (DOE), Office of Science, under Awards DE-SC0021385, DE-SC0013977, DE-SC0024425 (FRONTIERS SciDAC-5), and DE-SC0021113. This research used resources of the National Energy Research Scientific Computing Center (NERSC), and the Argonne and Oak Ridge Leadership Computing Facilities (ALCF and OLCF), DOE Office of Science User Facilities, supported under Contracts DE-AC02-05CH11231, DE-AC02-06CH11357, and DE-AC05-00OR22725, with allocations from ALCC and INCITE programs.
%\end{acknowledgments}

%\noindent \textit{Data availability. }
%The data that support the findings of this article are openly available \cite{Clauser_data}

%

%\funding{Sample text inserted for demonstration.}
% This section is a list of funder names and grant numbers

%\roles{Sample text inserted for demonstration.}
% List author names and the contributions made to the article, using terms from the NISO Contributor Roles Taxonomy (CRediT) https://credit.niso.org

\bibliographystyle{iopart-num}
\bibliography{NSTX-neg-density-mode}

%\end{multicols}

\end{document}